# Dual-beam intracavity optical trap with all-optical independent axial and radial self-feedback schemes

**Tengfang Kuang, Zijie Liu, Wei Xiong, Xiang Han, Guangzong Xiao[*], Xinlin Chen[**], Kaiyong Yang and Hui Luo**

*College of Advanced Interdisciplinary Studies, National University of Defense Technology, Changsha Hunan, 410073, China*

[*]*xiaoguangzong@nudt.edu.cn*

[**]*xlchencs@163.com*

**Abstract:** Recently single-beam intracavity optical tweezers have been reported and achieved orders-of-magnitude higher confinement than standard optical tweezers. However, there is only one feedback loop between the trapped particle's three-dimensional position and the scattering loss of the intracavity laser. That leads to the coupling effect between the particle's radial and axial motion, and aggravates the axial confinement efficiency. Here, we present and demonstrate the dual-beam intracavity optical trap enabling independent radial and axial self-feedback control of the trapped particle, through offsetting the foci of the clockwise and counter-clockwise beams. We have achieved the axial confinement efficiency of $1.6 \times 10^4$ mW$^{-1}$ experimentally at very low numerical aperture (NA=0.25), which is the highest axial confinement efficiency of the optical trap to date, to the best of our knowledge. The dual-beam intracavity optical trap will significantly expand the range of applications in the further studies of biology and physics, especially for a sample that is extremely sensitive to heat.

Since the invention of optical tweezers in 1986 by Arthur Ashkin *et al.* [1], optical traps have become an essential tool in several fields of physics [2,3], spectroscopy [4,5], biology [6,7], nanotechnology [8,9], and thermodynamics [10,11]. In these areas, higher optical confinement along the three axes has been desired to reduce the photothermal side-effect on the trapped particle [12]. We defined the optical confinement efficiency as the inverse positional variance per trapping power density on the trapped sample to characterize the confinement performance of the optical traps [13]. Many feedback control approaches have been used to improve system performance, including enhanced optical confinement efficiency and wider operational bandwidth [2,14,15]. However, the feedback control system applied to optical trapping usually requires complex hardware such as particle position sensing module, laser intensity modulator and electronic circuit, as well as signal processing procedures [16].

Kalantarifard *et al.* first presented and implemented self-feedback control of the particle's position in the single beam intracavity optical tweezers (SBIOT) [17]. They placed the particle inside the cavity of a ring fiber laser, as shown in Fig.1(a1). When the particle tries to escape from the trap region, it scatters less light so the optical loss decreases. The laser power resultantly increases and the particle is pulled back. The laser operation is nonlinearly coupled to the motion of the trapped particle. This coupling gives rise to intrinsic nonlinear feedback forces that confine microparticles efficiently at low intensity and low numerical apertures. Such systems have obtained a two orders of magnitude reduction of the average light intensity at the sample, compared with standard optical tweezers that achieve the same degree of confinement. Moreover, there is no requirement for any detection and control hardware or signal processing procedures compared with the standard optical tweezers with a feedback loop.

However, we noticed that the axial confinement efficiency is consistently about two orders of magnitude lower than the radial confinement efficiency [13]. The optical confinement efficiency is mainly determined by its own trapping stiffness of the trapping laser and the nonlinear feedback force in the SBIOT. Along the axial direction, the trapped particle is mainly balanced by the scattering force and the gravitational force, while the exerted gradient force is relatively weak. By contrast, the stronger radial gradient force creates the larger radial trapping stiffness and higher radial confinement efficiency. More importantly, there is only one feedback loop between the trapped particle's three-dimensional position and the scattering loss of the intracavity laser. That leads to the coupling effect between the particle's radial and axial motions, and aggravates the axial confinement efficiency.

In this Letter, we first present and demonstrate the dual-beam intracavity optical trap (DBIOT), which implements independent radial and axial self-feedback control of the trapped particle. The particle is trapped by the clockwise (CW) and counter-clockwise (CCW) beams inside the ring fiber laser. Through offsetting the foci of the CW and CCW beams, we create a separated self-feedback loop between their differential scattering loss and the axial position of the trapped particle. The nonlinear axial feedback force and resultant axial confinement efficiency can be adjusted by changing the offset between two foci of the CW and CCW beams. We achieve the axial confinement efficiency of $1.6 \times 10^4$ $mW^{-1}$ experimentally, which is about two orders of magnitude higher than the SBIOT in Ref [17].

Fig. 1 schematically shows the responses of the scattering loss and the intracavity laser power when a set of triangle axial displacements is applied to the particles trapped in SBIOT and DBIOT. The cylindrical coordinate systems were established as shown in Fig. 1(a). The cylindrical coordinate system is centered at the lens's focal point in Fig. 1(a1, a2). The origin of the cylindrical coordinate system in Fig. 1(a3) is located at the center between two foci of the trapping lenses. Compared with the SBIOT, we omit the optical isolator in the ring fiber cavity of the DBIOT as shown in Fig. 1(a). Both the CW and CCW beams are permitted to travel and construct a dual-beam optical trap in the ring cavity. The dependence of their scattering loss on the particle position is the same as that in the SBIOT [17]. Naturally, they will produce double radial nonlinear feedback optical force and generate stronger optical confinement. Next, we will focus on the axial trapping in SBIOT and DBIOT.

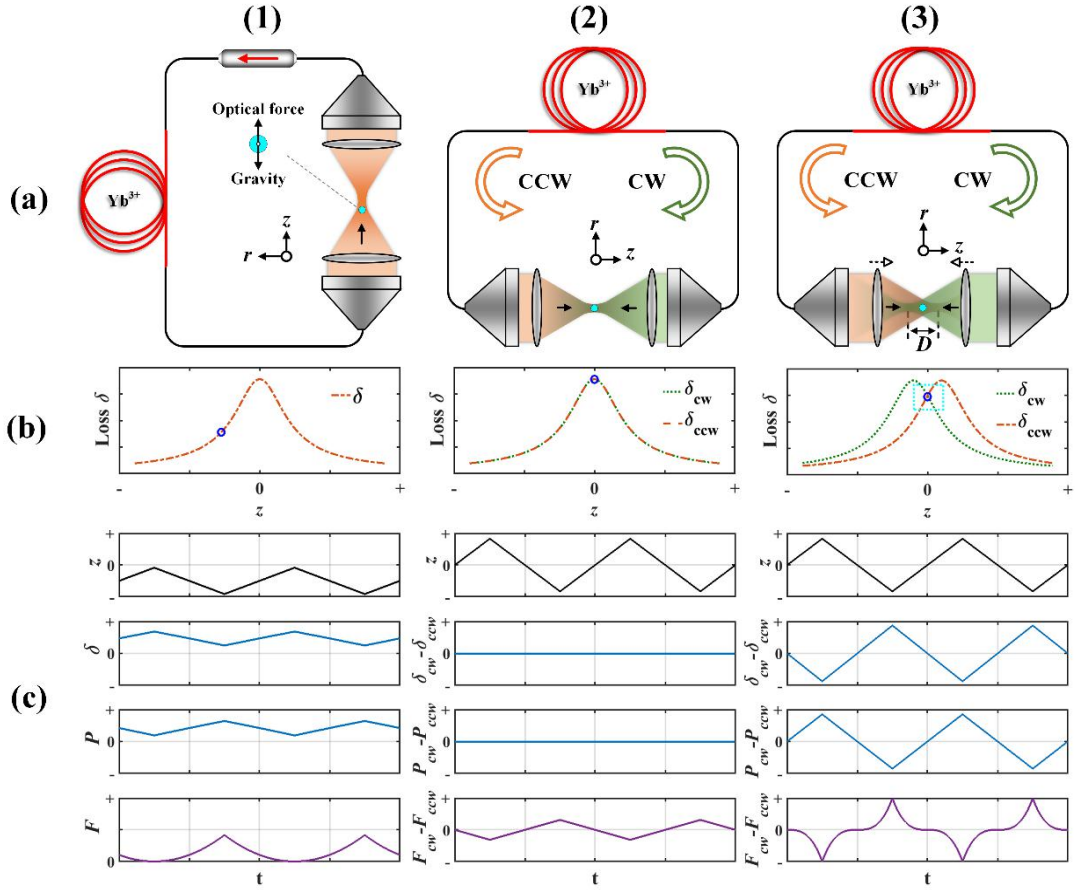

Fig. 1. Comparison of the characteristic between SBIOT (1) and DBIOT (2, 3). (a) The schematics of the SBIOT (a1) and DBIOT (a2, a3). The foci of two trapping lenses in (a3) is offset to a certain value *D*. (b) The cavity loss as a function of the particle's axial position. (c) The response of the intracavity optical powers and optical forces when a set of triangle axial displacement is applied on the trapped particle.

In the SBIOT, the coupling between the scattering loss $\delta$ and the axial motion of the trapped particle produces the axial nonlinear feedback force which dominates the axial optical confinement efficiency [18], as illustrated in Fig. 1(b1, c1). Note that the axial feedback force is constantly pushing the trapped particle away. The direction of the trapping beam must be directed against gravity. For the case of the DBIOT with no foci offset, the scattering losses of the CW and CCW laser respond nearly identically to the particle's axial motion and cancel out. So, the differential scattering loss of the CW and CCW lasers is independent of the particle's axial displacement (see Fig.1(b2, c2)) and no axial nonlinear feedback force is exerted on the trapped particle. As a result, only the weak axial scattering force gives rise to the low axial optical confinement efficiency. However, when we offset the foci of the CW and CCW beams in the DBIOT, the scattering loss curve of the two beams will deviate from each other and show inverse response to the axial displacement of the trapped particle as shown in Fig. 1(b3, c3). The differential scattering loss and the resultant differential laser intensity of the two beams will generate axial nonlinear inverse feedback optical force and improve the axial trapping stability.

Obviously the higher the change rates of the differential scattering loss with respect to the offset from the equilibrium position, $|d(\delta_{cw}-\delta_{ccw})/dz|$ and $|d\delta/dr|$, are, the larger the nonlinear inverse feedback optical forces are [18]. According to the physical model in Ref. [18], we calculate the $|d(\delta_{cw}-\delta_{ccw})/dz|$ and $|d\delta/dr|$ [19,20], as shown in Fig. 2. The numerical aperture of the trapping lens is set as NA=0.25. The wavelength of the trapped laser is 1030nm. The particle's radii are selected as 2, 3, 6, and 8 μm respectively. The result shows that the $|d(\delta_{cw}-\delta_{ccw})/dz|$ sharply increases first and then decreases when the foci offset rises. The $|d(\delta_{cw}-\delta_{ccw})/dz|$ reaches its peak around $D = 22$ μm. Orders-of-magnitude improvement of the $|d(\delta_{cw}-\delta_{ccw})/dz|$ is obtained compared with the case with no foci offset. Meanwhile, the $|d\delta/dr|$ is slightly reduced with the rise of the foci offset. Fig. 2 also shows that the stronger $|d(\delta_{cw}-\delta_{ccw})/dz|$ and $|d\delta/dr|$ are obtained for a larger particle.

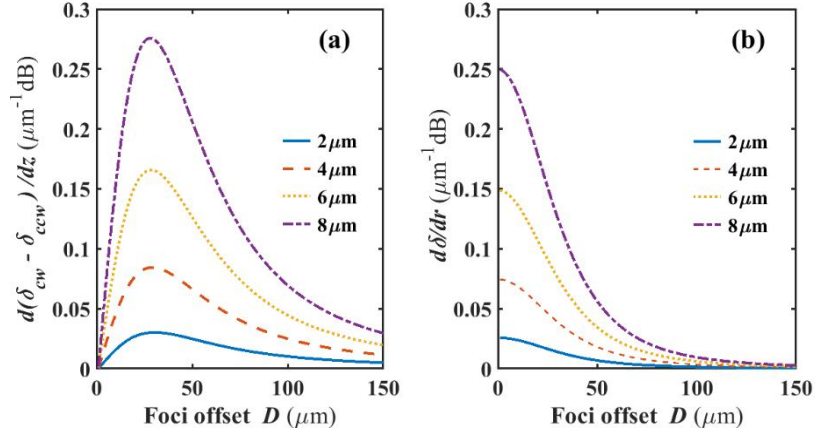

Fig. 2. Simulation results of the scattering loss. (a) The differential scattering loss per axial displacement and (b) the scattering loss per radial displacement versus the foci offset $D$.

The experimental setup is depicted in Fig. 3. It comprises a continuous-wave ring-cavity fiber laser emitting along the clockwise (CW) and counter-clockwise (CCW) directions. The Yb-doped gain medium (Yb 1200-6/125, nLIGHT, core diameter of 6 μm, cladding diameter of 125 μm) is pumped by a single-mode diode-laser at 976 nm through a wavelength division multiplexer $WDM_1$. The residual pumping laser is coupled out of the ring-cavity by another wavelength division multiplexer $WDM_2$, and terminates in a laser terminator. The CW and CCW lasers, centered at 1030nm, are expanded to free space by collimators (ZC618FC-B, Thorlabs) $C_1$, $C_2$ and coupled into the opposite fibers through the opposite collimators, implementing a ring-cavity. The lenses $L_1$ and $L_2$ (NA = 0.25) in the free space laser path compose a 1:1 beam expansion system. The particle is trapped by the focused CW and CCW lasers in the chamber. We use the dichroic mirror (1030 nm, R = 98 ± 0.2 %; 532 nm, R < 5 %) to reflect the majority of the intracavity laser. The 2% of the intracavity laser power along the collimator's direction is sent to the power meter (PDA50B2, Thorlabs) for power monitoring. The CW and CCW powers are detected by power meters $P_1$ and $P_2$, respectively. The 2% of the intracavity laser power along the trapping direction is sent to three balanced photo-detectors (PDB450C, Thorlabs) for detecting the three-dimensional position of the particle [2]. The CCD is used to observe the trapped particle.

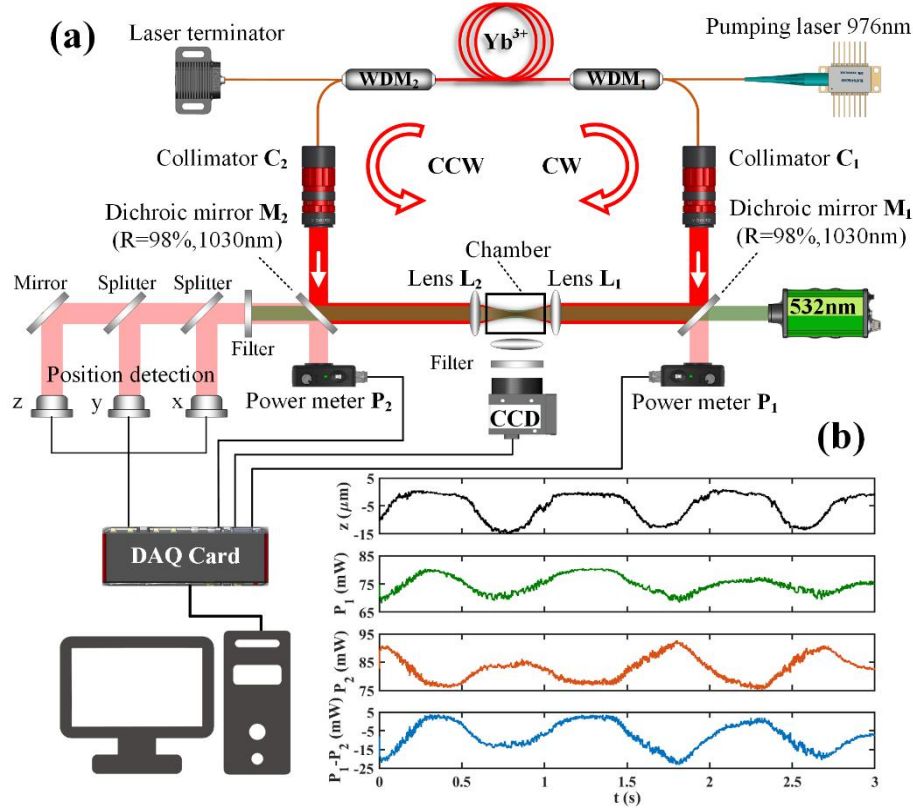

Fig. 3. Experimental setup. (a) The schematic of the experimental setup. Two red arrows represent the CW and CCW directions. WDM represents the wavelength division multiplexer. (b) The response of the intracavity optical powers when the particle is pushed by the 532 nm laser along the axial direction.

The pumping power is set as about 300 mW in the experiment. We use a collimated single-mode laser with the wavelength of 532 nm to push the particle along the axial direction. The light powers of the CW and CCW laser vary with the axial motion of the trapped particle as illustrated in Fig. 3(b). When the particle moves along the CW direction (-z direction), the CCW laser power increases and the CW laser power decreases. When the particle moves along the CCW direction (+z direction), the CCW laser power decreases and the CW laser power increases. The differential power of the CW and CCW laser ($P_{cw} - P_{ccw}$) shows the same trend as the particle's axial displacement, which is in great agreement with the prediction as illustrated in Fig. 1(c3).

We measured the optical confinement efficiency of the DBIOT for different foci offsets, as shown in Fig. 4. When the foci offset is $D = 0$ μm, the radial and axial confinement efficiencies are around $1.5 \times 10^3$ and $1.2 \times 10^4$ mW$^{-1}$ respectively. When the foci offset is increased, the axial confinement efficiency first rises and then decreases, and the radial confinement efficiency slightly declines. The axial confinement efficiency finds its optimal value of $1.6 \times 10^4$ mW$^{-1}$ when the foci offset is $D_{opt} = 24$ μm. The corresponding radial confinement efficiency reaches $8 \times 10^3$ mW$^{-1}$. This result demonstrates that adjusting the foci offset of the CW and CCW lasers will allow one to tailor the axial and radial nonlinear inverse feedback optical forces and regulate the axial and radial optical confinement efficiency of the DBIOT. That agrees well with the dependence of the scattering loss on the displacement of the particle presented in Fig.2.

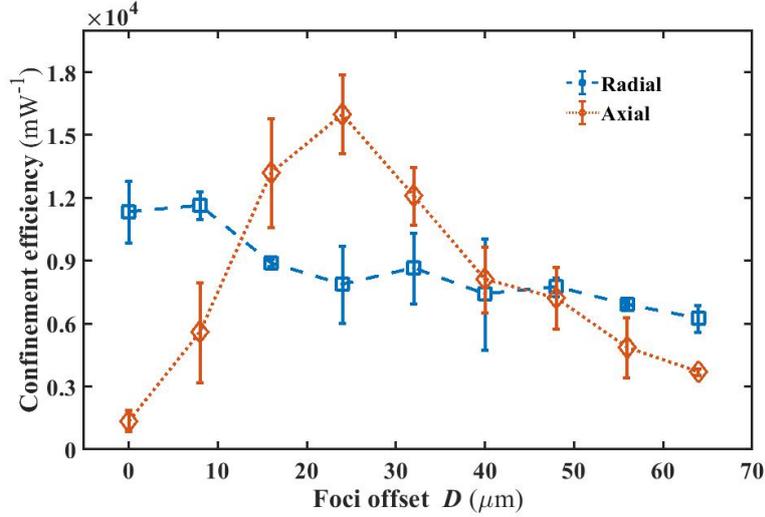

Fig. 4. Experimental results for the radial (□, blue) and axial (◊, red) optical confinement efficiency as a function of the foci offset $D$. The particle's diameter is 5 μm.

Fig 5 illustrates the comparison of the experimental confinement efficiency between DBIOT, SBIOT, standard optical tweezers (SOT) and standard dual-beam optical trap (SDBOT) for particles of various sizes. We set the foci offset of two beams as about 24 μm in the DBIOT where the $SiO_2$ sphere is trapped. Others included in this figure are compiled from published experiments [2,13,17]. It shows that, as a whole, the intracavity optical trap (i.e., DBIOT and SBIOT) realizes stronger optical confinement than the standard optical trap (i.e., SOT and SDBOT) due to the intrinsic self-feedback scheme. Moreover, the axial confinement efficiency achieved in the DBIOT is the highest among the above-mentioned optical traps, which is one order of magnitude higher than that for the SDBOT, and two orders of magnitude higher than that for the SBIOT. The improvement of the optical confinement efficiency is remarkable for the particle with larger diameter.

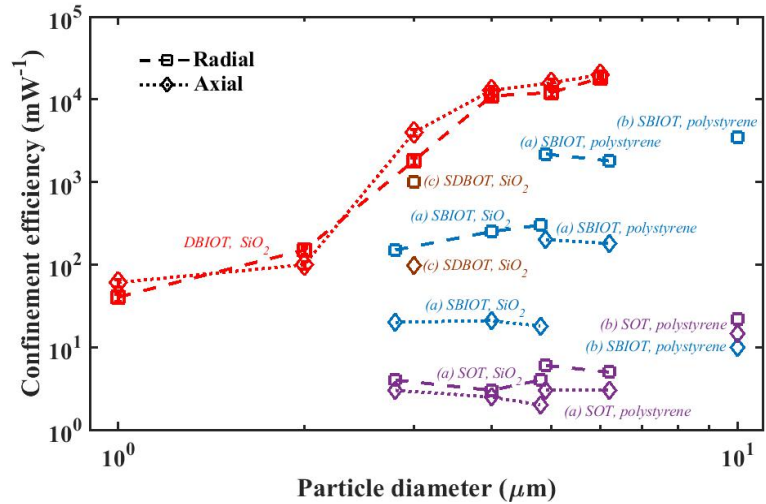

Fig. 5. Experimental results for the radial (□) and axial (◊) optical confinement efficiency of the dual beam intracavity optical trap (DBIOT, red, NA = 0.25), compared with single-beam intracavity optical tweezers (SBIOT, blue, (a) NA = 0.12, (b) NA = 0.25), standard optical tweezers (SOT, purple, NA = 1.25) and standard dual-beam optical trap (SDBOT, brown, NA = 0.68). The data except for the DBIOT is compiled from published experiments for published experiments: (a) ref. [17], (b) ref. [13], (c) ref. [2].

In conclusion, we first present and realize the dual-beam intracavity optical trap (DBIOT) enabling independent self-feedback control of the trapped particle's radial and axial positions. Besides the self-feedback loop between the scattering loss and the radial position of the trapped particle, we create a separated self-feedback loop between their differential scattering loss and the axial position through offsetting the foci of the CW and CCW beams. This feature is quite distinctive from the SBIOT in which the trapped particle's three-dimension displacements are fed back to one physical quantity i.e., the scattering loss of the intracavity laser [17]. Thus, the coupling effect between the particle's radial and axial motion will be reduced to a great extent [13]. As a result, the optical confinement efficiencies in the DBIOT are substantially enhanced by contrast with that in the SBIOT. Secondly, we achieve an adjustable axial and radial nonlinear inverse feedback optical forces in the DBIOT. This scheme allows tailoring both the axial and radial confinement efficiencies through tuning the foci offset of the CW and CCW beams. In our experiment, the axial confinement efficiency is dramatically improved to the same level as that along the radial direction. On contrast, the radial confinement is usually far more robust than the axial confinement in the dual-beam optical trap to date [21,22]. Finally, we have obtained the higher axial and radial confinement efficiencies in the DBIOT than all other state-of-the-art optical traps due to the nonlinear feedback force [17,18]. In particular, it should be noted that the direction of the trapping beam in the DBIOT is arbitrary, without the restriction due to gravity in the SBIOT [17].

Therefore, we expect that the DBIOT will enable higher three-dimension optical confinement than the SBIOT and thus pave the way for several research fields, especially in biology where the sample is extremely sensitive to light intensity [12], and in fundamental physical research where the thermal noise is harmful [23,24]. The DBIOT also may be a new cavity optomechanical system based on dissipative coupling. That is intrinsically different from the optomechanical system with levitated nano-particle based on the dispersive coupling [25,26]. We anticipate there will be many novel and interesting underlying physical phenomena when the DBIOT operates in air or vacuum.

**Funding**

National Natural Science Foundation of China (61975237, 11904405); Scientific Research Project of National University of Defense Technology (ZK20-14).

**Acknowledgment**

The authors wish to thank Bin Luo for the valuable suggestions.

**Disclosures**

The authors declare no conflicts of interest.